\documentclass[dvips]{jfm}
\usepackage[authoryear,round,sort]{natbib}
\usepackage{sublabel}
\usepackage{latexsym}
\usepackage[dvips]{color,graphicx}
\usepackage{amsmath}
\usepackage{pstricks}
\usepackage[dvips,colorlinks]{hyperref} 

\newpsobject{showgrid}{psgrid}{subgriddiv=1,griddots=10,gridlabels=6pt}

\newcommand {\norm} [1] { \lVert #1 \rVert}

\newcommand {\abs} [1] {\left| #1 \right|}

\numberwithin{equation}{section}

\title{Stable Manifolds and the Transition to Turbulence in Pipe Flow}

\author[
D. Viswanath
    and
P. Cvitanovi\'c
]{
$^1$D.\ns  V\ls I\ls S\ls W\ls A\ls N\ls A\ls T\ls H
\and
$^2$P.\ns C\ls V\ls I\ls T\ls A\ls N\ls O\ls V\ls I\ls \'C
}

\affiliation{
$^1$Department of Mathematics,
University of Michigan,
Ann Arbor, MI 48109, USA\\
[\affilskip]
$^2$School of Physics,
Georgia Institute of Technology,
Atlanta, GA  30332, USA
}

\begin{document}
\maketitle

\begin{abstract}
Lower-branch traveling waves and equilibria computed in pipe flow and
other shear flows appear intermediate between turbulent and laminar
motions. We take a step towards connecting these
lower-branch solutions to transition by deriving a numerical method
for finding certain special disturbances of the laminar flow in a
short pipe. These special disturbances cause the disturbed
velocity field to approach the lower-branch solution by evolving along
its stable manifold. If the disturbance were slightly smaller, the
flow would relaminarize, and if slightly larger, it would transition
to a turbulent state.
\end{abstract}

\section{Introduction}

The connection between the law of resistance to water flowing in a
tube and the sinuous or direct nature of the internal motion of the
fluid was established by \cite{Reynolds}. In the transitional regime,
he observed ``flashes'' and recorded them in Figure 16 of his paper.
Later experiments using hot-wire measurements have revealed the
structure of puffs and slugs
\citep{WC}. Particularly intriguing are equilibrium puffs that
maintain their spatial extent as they travel downstream with a
characteristic speed \citep{WSF}. Such puffs are approximately
$20$ pipe diameters long
and are observed for $Re$ (Reynolds number) somewhat
greater than $2000$. Further, the structure of the puff is independent
of the disturbance used to create it.

For a range of $Re$, the flow injected into the pipe assumes the
familiar Hagen-Poiseuille laminar profile downstream. The flow does
not have the laminar profile at the inlet. Therefore it is important
to distinguish between disturbances at the inlet and disturbances to
fully developed laminar flow
\citep{WPKM08}[{\it Willis, Peixinho et al.}]. The early experiments used inlet disturbances, but in
theoretical investigations such as this one, it has been common
practice to consider disturbances to the laminar flow.

In their experiments to determine the dependence of the threshold for
transition on $Re$, \cite{DM} used a constant mass-flux pipe and
introduced disturbances to the laminar flow at a point sufficiently
downstream from the inlet.  They were able to determine thresholds
over a range of $Re$, but Figure 11 and other figures in their paper
showed that certain disturbances above the threshold do not transition
while certain disturbances below the threshold do
transition. \cite{HJM} determined that the threshold scaled as
$Re^{\alpha}$ with $\alpha = -1$, when the laminar flow was disturbed
by a single boxcar pulse of fluid injected at six different points.
\cite{MM2} have reproduced $\alpha = -1$ in a numerical study that
added a body force term to the Navier-Stokes equation to model the
effect of the boxcar pulse of fluid.  For different disturbances of
the laminar flow,
\cite{PexMull07} found  $\alpha < -1$.
In that experiment, the transition is sequential, with flow
visualizations showing the disturbance changing its form as it travels
downstream before leading to bigger structures.  Following \cite{OB},
\cite{PexMull07} point out that disturbances that lead to $\alpha < -1$
probably do not significantly distort the mean flow.

With regard to theory, Reynolds's
assertion that ``there was small chance of discovering anything new or
faulty'' in the Navier-Stokes equation has stood the test of time.
Thanks to numerical computations, we now know that the incompressible
Navier-Stokes equation adequately explains a remarkable wealth of
phenomena related to transitional turbulence and fully developed
turbulence.  Although the Navier-Stokes equation can be solved
numerically in certain regimes, the nature of the solutions of that
equation has proved difficult to understand. 

It is clear, however, that the nature of the solutions is quite different
in the turbulent and transitional regimes. Fully developed turbulence
is characterized by rapid decay of correlations and fine
scales. Statistical theories that separate turbulent velocity fields
into means and fluctuations have had significant successes
\citep{Narasimha}, even though coherent motions are present in certain
regions of fully developed turbulence
\citep{Robinson}. In contrast, the transition problem seems to be
fundamentally dynamical in nature.

Following the experiments of \cite{DM}, \cite{FE1} and \cite{SE}
argued that there is no sharp boundary between initial conditions that
trigger turbulence and those that do not, and demonstrated
computationally that the stability border for plane Couette flow is a
fractal.  One of their suggestions, namely that a chaotic saddle could be
present for transitional $Re$, illustrates the dynamical nature
of the transition problem.

More recently, \cite{FE2} and \cite{WK} computed a number of traveling
wave solutions of pipe flow. Their work was preceded by computations
of somewhat similar solutions of channel flows by
\cite{Nagata1} and \cite{Waleffe1}. \cite{HD} found streak patterns
in puffs and slugs that appeared  close to those of some pipe
flow traveling waves. The traveling waves were computed in short
pipes, typically only a few pipe diameters long, while puffs are as
long as $20$ pipe diameters. Therefore the following question may be
asked: do the experimentally observed structures correspond to
the computed traveling waves?

The correlation functions, such as those of \cite{SchEckVoll07},
used to detect
streak patterns in experimental or numerical flow fields look for
$m$-fold rotational symmetry with respect to the pipe axis. Using such
a correlation function, Figure 5 of \cite{SchEckVoll07} illustrates a
transition from a four streak state to a six streak state within a
spatial range of a single pipe radius. \cite{WillKer08} found
structures with $m=3$ and $m=4$ within streamwise distances of about
$2$ pipe diameter preceding the trailing edge of the puff and $5$ pipe
diameters following the trailing edge. Figure 5 of their paper gives
some evidence that parts of the puff on either side of its
trailing edge (but not at the trailing edge itself)
visit traveling wave solutions with $m=3$ and $m=4$.  It
could be significant that Figures 7 and 23 of \cite{WK} (which use
different units) imply that some of the $m=3$ and $m=4$ traveling
waves have wave speeds relatively close to that of the
puff. \cite{WillKer08} also found that the qualitative comparisons of
\citep{HD} which used slug cross-sections had significant
problems. Instead of the trailing edge of the puff, \cite{ES} used the
center of turbulent energy to fix a position within a moving
puff. Their center of turbulent energy is more precisely defined and
it moves with the puff in a quite regular manner. They found the axial
correlation lengths around that center to be quite short. While the
question raised in the previous paragraph cannot be answered
conclusively at the moment, these arguments suggest that short pipe
computations are of some relevance.

Another point to be mentioned is that puffs, in which streak patterns
resembling those of some traveling waves have been detected, are
observed experimentally only for $Re<2800$. The transition experiments
that measure thresholds \citep{HJM, PexMull07} reach $Re$ as high as
$20000$.  Thus the relevance of puffs to transition may seem limited.
However, there is a possibility that puffs exist as solutions of the
Navier-Stokes equation beyond the $Re$ at which they are observed in
experiments \citep{WillKer09}.

Most of the lower branch solutions of pipe flow and the channel flows
seem to be on the laminar-turbulent boundary
\citep{duguet07, GHC, IT, Kawahara, Kerswell, KT, SGLDE08, DVCouette2, WGW},
which means that for some tiny disturbances of the lower branch
solution, the disturbed state evolves and becomes laminar
uneventfully.  For other disturbances, the disturbed stated evolves
and becomes turbulent, or undergoes a turbulent episode before it
becomes laminar.

In this article, we investigate if there are small disturbances of
the laminar solution, for which the disturbed state evolves and hits a
given lower branch solution. By small, we mean firstly that the
magnitude of the disturbance should decrease algebraically with $Re$
and secondly that the disturbance should not change the mean flow
significantly. The existence of such a disturbance would establish
that the flow can transition from laminar to turbulence by passing
through the vicinity of the given lower branch traveling wave.

In fact, \cite{KLH} found such disturbances when computing thresholds
without fully realizing that they were hitting a lower branch
equilibrium solution of plane Couette flow. But the situation they
tackled is an especially simple one because the lower branch solution
has a single unstable direction \citep{SGLDE08, TI, DVCouette2, WGW}. We
consider the asymmetric traveling wave computed by \cite{PK}.
That traveling wave has two unstable directions both of which lie in a
symmetric subspace, and thus serves to illustrate that the method used
for computing thresholds cannot be used to hit traveling waves that
have more than one unstable direction.

\begin{figure}
\begin{center}
\includegraphics[scale=0.4]{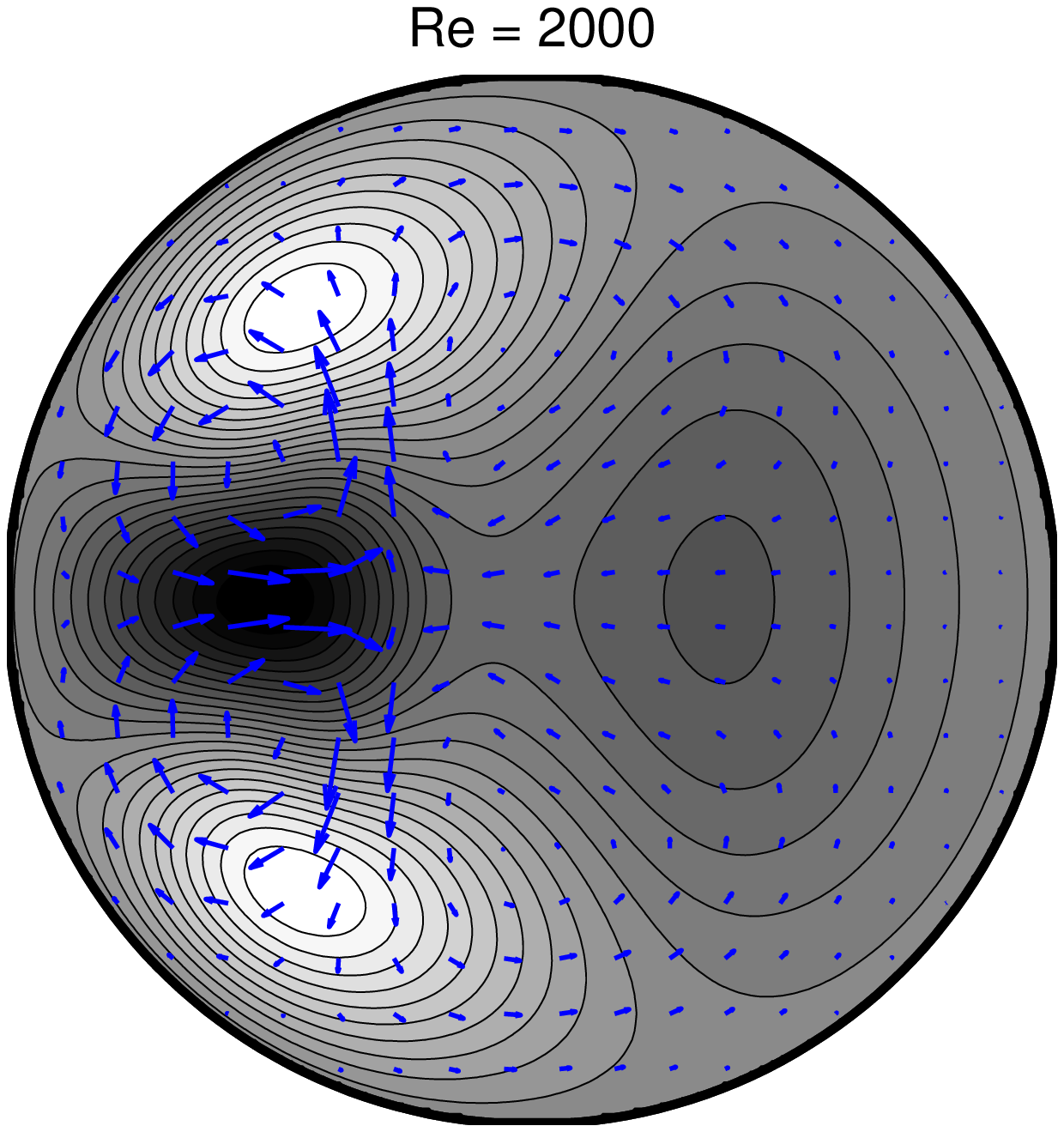}
\hspace*{1cm}
\includegraphics[scale=0.4]{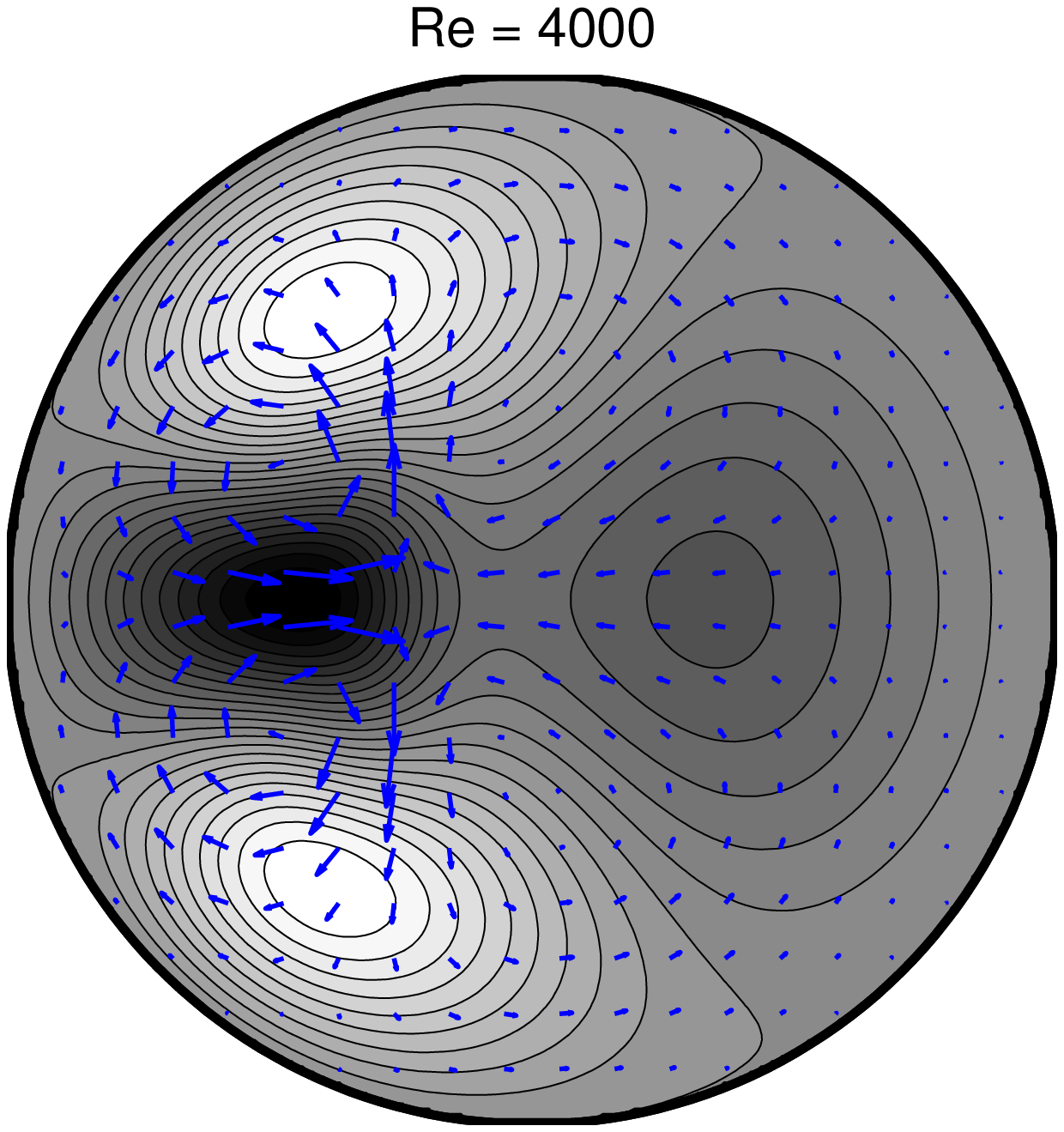}
\end{center}
\caption[xyz]{Contour plots of $z$-averaged streamwise velocity with the
laminar flow subtracted. Rolls are superposed. The contour levels are
equispaced in the intervals $[-0.195,0.170]$ and $[-0.191,0.148]$,
respectively, with the lighter regions being faster. The maximum
magnitude of the vectors in superposed quiver plots are $0.0075$ and
$0.0038$, respectively.}
\label{fig-1}
\end{figure}

\begin{table}
\begin{center}
\begin{tabular}{c|ccccccc}
$Re$ & $I=D$ & $ke$ & $ke_d$ & $ke_0$ & $ke_1$ & $ke_2$ & $ke_3$ \\
\hline
$2000$ & $1.0881$ & $0.9783$ & $0.013$ & $0.9778$ &
$4.4e\!\!-\!\!4$ & $2.3e\!\!-\!\!5$ & $2.2e\!\!-\!\!7$ \\ 
$2500$ & $1.0802$ & $0.9790$ & $0.012$ & $0.9788$ & $2.8e\!\!-\!\!4$ &
$1.2e\!\!-\!\!5$ & $1.0e\!\!-\!\!7$ \\ 
$3000$ & $1.0755$ &
$0.9794$ & $0.012$ & $0.9792$ & $1.9e\!\!-\!\!4$ & $7.6e\!\!-\!\!6$ &
$5.9e\!\!-\!\!8$ \\ 
$4000$ & $1.0705$ & $0.9796$ & $0.012$ &
$0.9795$ & $1.1e\!\!-\!\!4$ & $3.8e\!\!-\!\!6$ & $2.8e\!\!-\!\!8$
\end{tabular}
\end{center}
\caption[xyz]{
The kinetic energy of the traveling wave and the kinetic
energy with laminar flow subtracted are denoted by $ke$ and $ke_d$,
respectively. The last
four columns give the kinetic energy in modes with $n=0, \pm 1, \pm 2, \pm 3$.}
\label{table-1}
\end{table}

The asymmetric traveling wave of \cite{PK}, which has two fast streaks
located near one side of the pipe, is shown in Figure
\ref{fig-1}. The preferential location of the streaks towards one
side is also found in edge states that occur in transition
computations \citep{SEY}. Table \ref{table-1} gives basic data for
that traveling wave at four different $Re$. That data will be useful
for judging the closeness of approaches to the traveling wave. The
choice of units, the significance of $I$ and $D$ in Table
\ref{table-1}, and the meaning of the streamwise modes with $n=0, \pm
1, \pm 2,
\pm3$ are explained in Section 2. More extensive data for the asymmetric
traveling wave can be found elsewhere \citep{DVPipe1}.

To find a small disturbance of the laminar solution that evolves into
a given lower-branch state, it is necessary to consider a linear
superposition of disturbances whose dimension equals that of the
unstable manifold of the lower-branch state. That requirement follows
from a consideration of the co-dimension of the stable manifold of the
lower-branch state. In addition, the disturbances that are linearly
combined must be chosen carefully.

To hit the asymmetric traveling wave, we consider three different
disturbances of the laminar solution. The first disturbance is
obtained by extracting the rolls, which are formed by averaging the
traveling wave in the streamwise direction and retaining only the
radial and azimuthal components of the velocity field.
The choice of rolls is related to the so-called lift-up
mechanism \citep{Landahl}.
The two other
disturbances are the two unstable eigenvectors of the traveling wave. If
one of the eigenvectors is added to the traveling wave, it has the
effect of either reinforcing or weakening the fast streaks. The other
eigenvector seems to alter the location of the fast streaks. It must be
remembered, however, that the disturbances are added to the laminar
solution and not to the traveling wave.

In Section 4, we show that disturbances of the laminar flow obtained
by varying any two of these three disturbances evolve and hit the
traveling wave. The choice of the unstable eigenvectors might seem puzzling
as our intention is to hit the traveling wave and not to move away
from it. The reason that choice works is partially explained in
Sections 3 and 4. Section 4 also shows that the magnitudes of
the disturbances needed to hit the asymmetric traveling wave diminish
algebraically with $Re$.

All our computations use a pipe that is $\pi$ pipe diameters long and
the traveling waves are computed with $85715$ active degrees of
freedom. The computations of relative periodic solutions (or modulated
traveling waves) in plane Couette flow use triple the number of
degrees of freedom \citep{DVCouette1}, although those computations are
roughly $5$ to $10$ times as expensive with the same number of degrees
of freedom. The computation of traveling waves is an insignificant
part of the total computational expense, however, as will become clear
in Section 4.  We need to use a short pipe to keep the total
computational expense manageable.

The pipe we use is too short to capture transitional structures such
as puffs. To add to the earlier discussion of the relevance of short
pipe computations of traveling waves, we mention the work of
\cite{MM1} which seems to suggest that transition scenarios can be
independent of pipe length.  The logic which is used to find
disturbances of the laminar solution that hit the
asymmetric traveling wave makes fairly intricate use of the dynamical
properties of the traveling wave. More work is needed to determine if
the same logic is applicable to transition in pipes of more realistic
length.

\section{Preliminaries}

The code for direct numerical simulation of pipe flow uses
cylindrical coordinates with $u$, $v$, and $w$ being the components of
the velocity in the radial ($r$), polar ($\theta$), and axial ($z$)
directions, respectively. The boundary conditions are no-slip at the
walls and periodic in the $z$ direction with constant mass-flux.  The
length of the periodic domain in the $z$ direction is denoted by $2\pi
\Lambda$. We use $\Lambda = 1$ throughout.
The units for distance and velocity are chosen so that
the pipe radius is $1$ and the Hagen-Poiseuille profile is given by $w
= 1 - r^2$.  The Reynolds number $Re$ is based on the pipe radius,
centerline velocity of the Hagen-Poiseuille flow, and kinematic
viscosity $\nu$.  The unit of mass is chosen so that the density of
the fluid is $1$. The units and boundary conditions follow those of
\citet{FE1}.

Let $\bar{w}(r)$ denote the mean velocity in the axial direction,
and $\bar{v}(r)$  the mean velocity in the polar direction. The
mass-flux per unit area is given by $2\int_0^1 r\bar{w}(r)\,dr$ and is
equal to $1/2$ for all velocity fields that obey the boundary
condition. The pressure gradient necessary to maintain constant
mass-flux varies from instant to instant. For the Hagen-Poiseuille flow,
it is $-4/Re$.

The spatial discretization is spectral. The radial component of the
velocity $u$ is represented as
\begin{equation}
u(r,\theta, z) = \sum_{n=-N}^{n=N}\sum_{m=-M}^{m=M}
\hat{u}_{n,m}(r) \exp(i m \theta) \exp(i n z/\Lambda).
\label{eqn-2-1}
\end{equation}
For the velocity field to be regular at $r=0$, the coefficients
$\hat{u}_{n,m}(r)$ must be even functions of $r$ for odd $m$ and odd
functions of $r$ for even $m$. Thus the functions can be reconstructed
by storing their values at $r = \cos(i\pi/L)$, $i = 0, 1, \ldots,
(L-1)/2$.  We assume $L$ odd so that there is no point at $r=0$
\citep{Trefethen}.  The radial component of vorticity is denoted by
$\xi$. It is represented in the same way $u$ is represented. The other
quantities used to represent the velocity field are $\bar{v}(r)$,
which is an odd function of $r$, and $\bar{w}(r)$ which is an even
function of $r$. The velocity field is constructed using $u, \xi,
\bar{v},
\bar{w}$ and the divergence free condition.  The advection term was
dealiased using the Orszag $3/2$ rule. All the computations
use $(N,M,L)=(16,18,81)$.

The rate of energy dissipation per unit mass is given by
$2 D/Re$, where $D$ is the integral of
\begin{equation}
\frac{1}{4 \pi^2\Lambda}
\biggl(
\frac{1}{r^2}\bigl(u^2+v^2 - 2 \frac{\partial u}{\partial \theta} v
+ 2u \frac{\partial v}{\partial \theta}\bigr) +
\sum_{U=u,v,w} \biggl(\frac{\partial U}{\partial r}\biggr)^2
+  \biggl(\frac{\partial U}{\partial z}\biggr)^2
+  \frac{1}{r^2}\biggl(\frac{\partial U}{\partial \theta}\biggr)^2
\biggr)
\label{eqn-2-2}
\end{equation}
over the volume of the pipe.  In its more familiar form, $D$ is the
integral of the sum of the norms of gradients of the three components
of the velocity field \citep{WK}.  The term under the summation in
\eqref{eqn-2-2} gives $\abs{\nabla{U}}^2$ for a scalar field $U(r,\theta,z)$.
The terms outside the summation in
\eqref{eqn-2-2} arise as cross terms when that operator is applied to
$u \cos\theta - v \sin\theta$ and $u\sin\theta + v\cos\theta$.  The
explicit form of \eqref{eqn-2-2} displays the $1/r^2$ singularities
that are hidden in vector notation.  Because those singularities
cancel at $r=0$, the numerical evaluation of $D$ in a spectral code is
a delicate matter. The rate of energy input per unit mass is given by
$2 I/Re$, where
\begin{equation}
I = -\frac{Re}{4\pi^2\Lambda}\int \nabla\cdot(p{\bf u}),
\label{eqn-2-3}
\end{equation}
with $p$ being pressure and with the integral being over the volume
of the pipe. For the Hagen-Poiseuille laminar flow, both $D$ and $I$
evaluate to $1$.

Figure \ref{fig-1} shows the asymmetric traveling wave solution first
computed by
\cite{PK}. To compute that traveling wave, we added rolls which approximate
the pattern in Figure \ref{fig-1} to the laminar solution and evolved
the velocity field to allow the streaks to develop. The resulting
velocity field was used as the initial guess for the GMRES-hookstep
method, which converged without a hitch.  The number of active degrees
of freedom in the representation of a velocity field is
$(L-2)+((2N-1)(2M-1)-1)(L-3)/2$.  The method uses translation
operators to handle the invariance of the pipe-flow equation with
respect to shifts along $z$ and rotations along $\theta$. These
operators are given by
\begin{align}
\mathcal{T}_1 u(r,\theta, z) &=
\sum_{m,n} i m \hat{u}_{n,m}(r)\exp(im\theta)\exp(inz/\Lambda)\nonumber\\
\mathcal{T}_2 u(r,\theta, z) &=
\sum_{m,n} (in/\Lambda)  \hat{u}_{n,m}(r)\exp(im\theta)\exp(inz/\Lambda),
\label{eqn-2-4}
\end{align}
where the indices $m,n$ correspond to the representation \eqref{eqn-2-1}.
A detailed description of the GMRES-hookstep method can be found
elsewhere \citep{DVCouette1, DVPipe1}.

The equations of pipe flow are unchanged by the  shift-reflect
symmetry:
\begin{align}
u(r, \theta, z) &\rightarrow u(r, -\theta, z+\pi\Lambda)\nonumber\\
v(r, \theta, z) &\rightarrow -v(r, -\theta, z+\pi\Lambda)\nonumber\\
w(r, \theta, z) &\rightarrow w(r, -\theta, z+\pi\Lambda).
\label{eqn-2-5}
\end{align}
The velocity field of the traveling wave of Figure \ref{fig-1} is
also unchanged by this discrete symmetry.

The magnitudes of disturbances and the norms of velocity fields are
given in Section 4 and other places using the square root of kinetic
energy norm. The kinetic energy, which is reported in tables such as
Table \ref{table-1}, is normalized to be $1$ for laminar flow.

To conclude this section, we mention a technical point about pipe flow
simulation using spectral codes that appears not to have been
discussed in the literature.  Once the advection term is computed, the
equations for evolving the modes decouple for pairs $(m,n)$ such that
the resulting equations depend only upon $r$ for a fixed $(m,n)$. The
decoupled equations will have terms with the factor $m^2/r^2 +
n^2/\Lambda^2$ in the denominator, and because of that factor the
terms will have a singularity at the point $r = -i m\Lambda/n$ in the
complex plane. If the number $2N$ of grid points in the streamwise
direction is increased while keeping the pipe length $2\pi\Lambda$
fixed, that singularity moves closer to the real line with greater
values of $n$ now being allowed.  When the singularity moves closer to
the real line, one has to use more grid points in the $r$ direction
to solve the decoupled equations with the same level of accuracy
\citep{Trefethen}.

\section{Unstable manifold of the traveling wave}

\begin{table}
\begin{center}
\begin{tabular}{c|ccc}
$Re$ & $\lambda_1$ & $\lambda_2$ & $\lambda_{3}$ \\ \hline
$2000$~ & ~$0.03247$ & $0.00897$ & $-0.00594$  \\ 
$2500$~ & ~$0.03049$ & $0.00725$ & $-0.02282 +i0.02041$  \\ 
$3000$~ & ~$0.02861$ & $0.00631$ & $-0.01978 +i0.01664$  \\ 
$4000$~ & ~$0.02529$ & $0.00531$ & $-0.01536 +i0.01190$
\end{tabular}
\end{center}
\caption[xyz]{
$\lambda_1$ and $\lambda_2$ are
the only unstable eigenvalues. $\lambda_3$ has the
greatest real part among the stable eigenvalues whose
eigenvectors lie in the shift-reflect invariant subspace.
}
\label{table-2}
\end{table}

To find disturbances of the laminar solution that evolve and hit the
asymmetric traveling wave, it is essential to understand the unstable
directions and the unstable manifold of that traveling wave.  Suppose
we disturb the laminar solution using rolls of the appropriate form
and some ``noise'', the magnitude of which is a fixed fraction of that
of the rolls, to introduce streamwise dependence. The disturbed state
will evolve and develop streaks. At the point of closest approach to
the traveling wave, we can think of the evolving velocity field as the
traveling wave plus two components, one of which is a combination of
the stable eigenvectors of the traveling wave with the other being a
combination of the unstable eigenvectors.  The stable eigenvectors
will decay under evolution. However, the component along the
unstable eigenvectors will be amplified and will take the evolving velocity
field away from the traveling wave. To ensure that the disturbed state
hits the traveling wave, the disturbance has to be arranged in such a
way that the evolving velocity field is free of the unstable
directions as it approaches the traveling wave.

Such a disturbance is easiest to arrange, if the traveling wave has
only one unstable direction. The component along that direction at the
point of closest approach can be eliminated by simply varying the
magnitude of the initial disturbance. However, the asymmetric
traveling wave has two unstable directions as shown in Table
\ref{table-2}. The two unstable eigenvalues $\lambda_1$ and
$\lambda_2$ decrease with increasing $Re$ at rates given by
$Re^{-0.41}$ and $Re^{-0.87}$, respectively \citep{DVPipe1}. Table
\ref{table-2} also shows the leading stable eigenvalue.

\begin{figure}
\begin{center}
\includegraphics[scale=0.35]{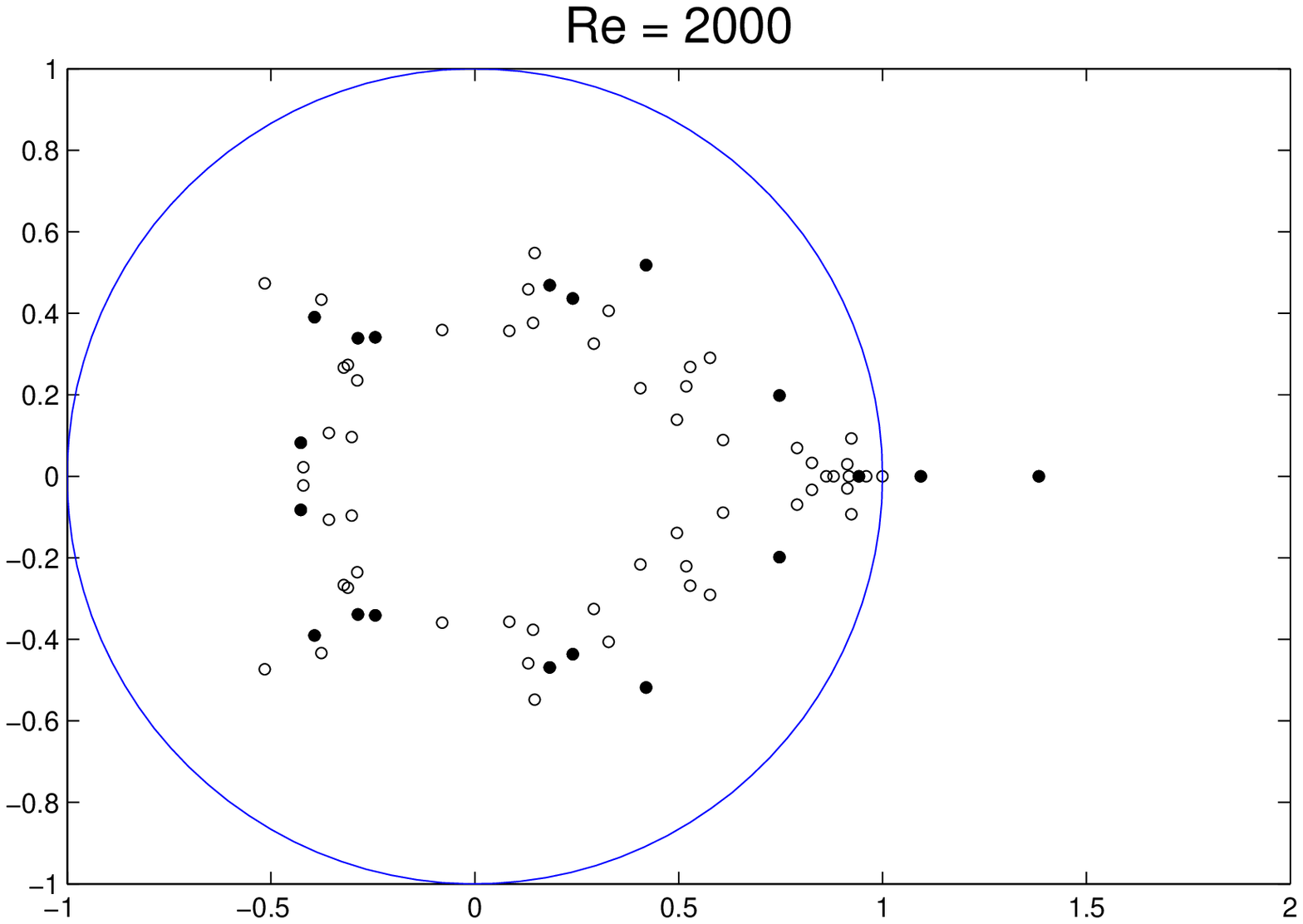}
\hspace*{1cm}
\includegraphics[scale=0.35]{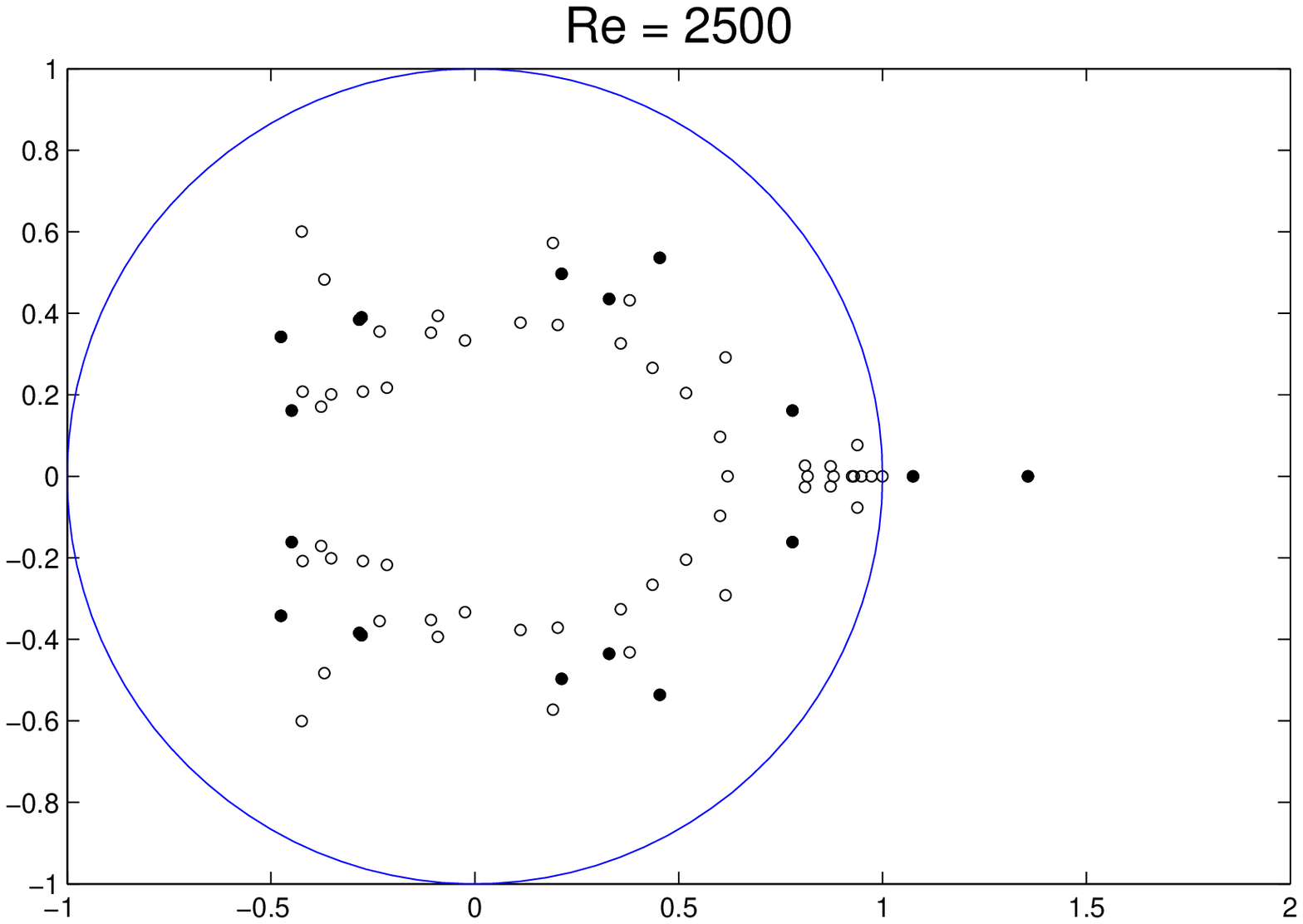}
\end{center}
\caption[xyz]{For eigenvalues $\lambda$ of the traveling wave, the
plots show $\exp(10\lambda)$ as circles for easier visualization.
If the corresponding eigenvector lies in the shift-reflect invariant
subspace, the circle is solid.}
\label{fig-2}
\end{figure}

Figure \ref{fig-2} gives a more complete idea of the spectrum of the
linearization around the asymmetric traveling wave. The spectra at
different $Re$ were computed using the Arnoldi iteration.  Some of the
interior eigenvalues near the centers of the circles in Figure
\ref{fig-2} are omitted. But we are certain that no unstable
eigenvalues are omitted. In addition, we have verified that none of the
eigenvalues in the figure is spurious.

\begin{table}
\begin{center}
\begin{tabular}{c|cccc|cccc}
 & \multicolumn{4}{c|}{$\lambda_1$} & \multicolumn{4}{c}{$\lambda_2$} \\ 
$Re$ & $ke_0$ &  $ke_1$ &  $ke_2$ &  $ke_3$ &  $ke_0$ &  $ke_1$ &  $ke_2$ &  $ke_2$ \\ \hline
$2000$ & $6.9e\!\!-\!\!1$ &  $2.6e\!\!-\!\!1$ &  $4.9e\!\!-\!\!2$ &  $1.1e\!\!-\!\!3$ &  $1.2e\!\!-\!\!1$ &  $7.4e\!\!-\!\!1$ &  $1.4e\!\!-\!\!1$ &  $3.1e\!\!-\!\!3$ \\ 
$2500$ & $7.0e\!\!-\!\!1$ &  $2.6e\!\!-\!\!1$ &  $4.2e\!\!-\!\!2$ &  $7.6e\!\!-\!\!4$ &  $1.1e\!\!-\!\!1$ &  $7.7e\!\!-\!\!1$ &  $1.2e\!\!-\!\!1$ &  $2.3e\!\!-\!\!3$ \\ 
$3000$ & $7.0e\!\!-\!\!1$ &  $2.6e\!\!-\!\!1$ &  $3.8e\!\!-\!\!2$ &  $6.2e\!\!-\!\!4$ &  $1.1e\!\!-\!\!1$ &  $7.8e\!\!-\!\!1$ &  $1.1e\!\!-\!\!1$ &  $1.9e\!\!-\!\!3$ \\ 
$4000$ & $7.0e\!\!-\!\!1$ &  $2.6e\!\!-\!\!1$ &  $3.2e\!\!-\!\!2$ &  $4.9e\!\!-\!\!4$ &  $1.1e\!\!-\!\!1$ &  $7.9e\!\!-\!\!1$ &  $9.4e\!\!-\!\!2$ &  $1.5e\!\!-\!\!3$
\end{tabular}
\end{center}
\caption[xyz]{
The kinetic energies in the $n=0,\pm 1,\pm 2,\pm 3$ modes of
the $\lambda_1$ and $\lambda_2$ eigenvectors. The eigenvectors are
normalized to have kinetic energy equal to $1$. }
\label{table-3}
\end{table}

Data for the unstable eigenvectors is given in Table \ref{table-3}.
Most of the
kinetic energy of the traveling waves themselves is in the $n=0$ (or mean)
mode, as shown in Table \ref{table-1}. Much of the kinetic energy
remains in $n=0$ for the $\lambda_1$ eigenvector, although $n=1$ now
has more than a quarter of the kinetic energy. For the $\lambda_2$
eigenvector, the $n=1$ mode dominates.

\begin{figure}
\begin{center}
\includegraphics[scale=0.4]{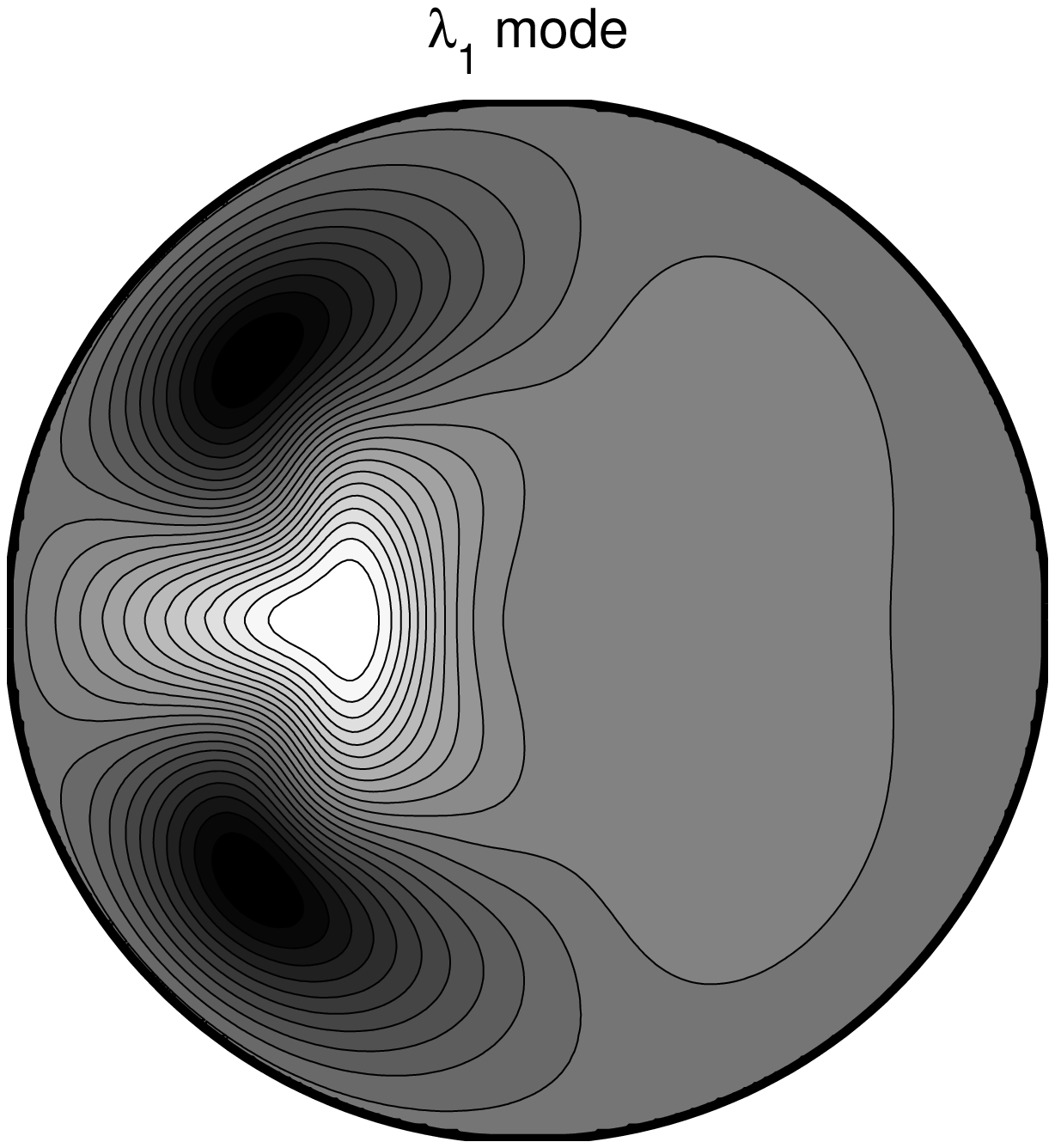}
\hspace*{.5cm}
\includegraphics[scale=0.4]{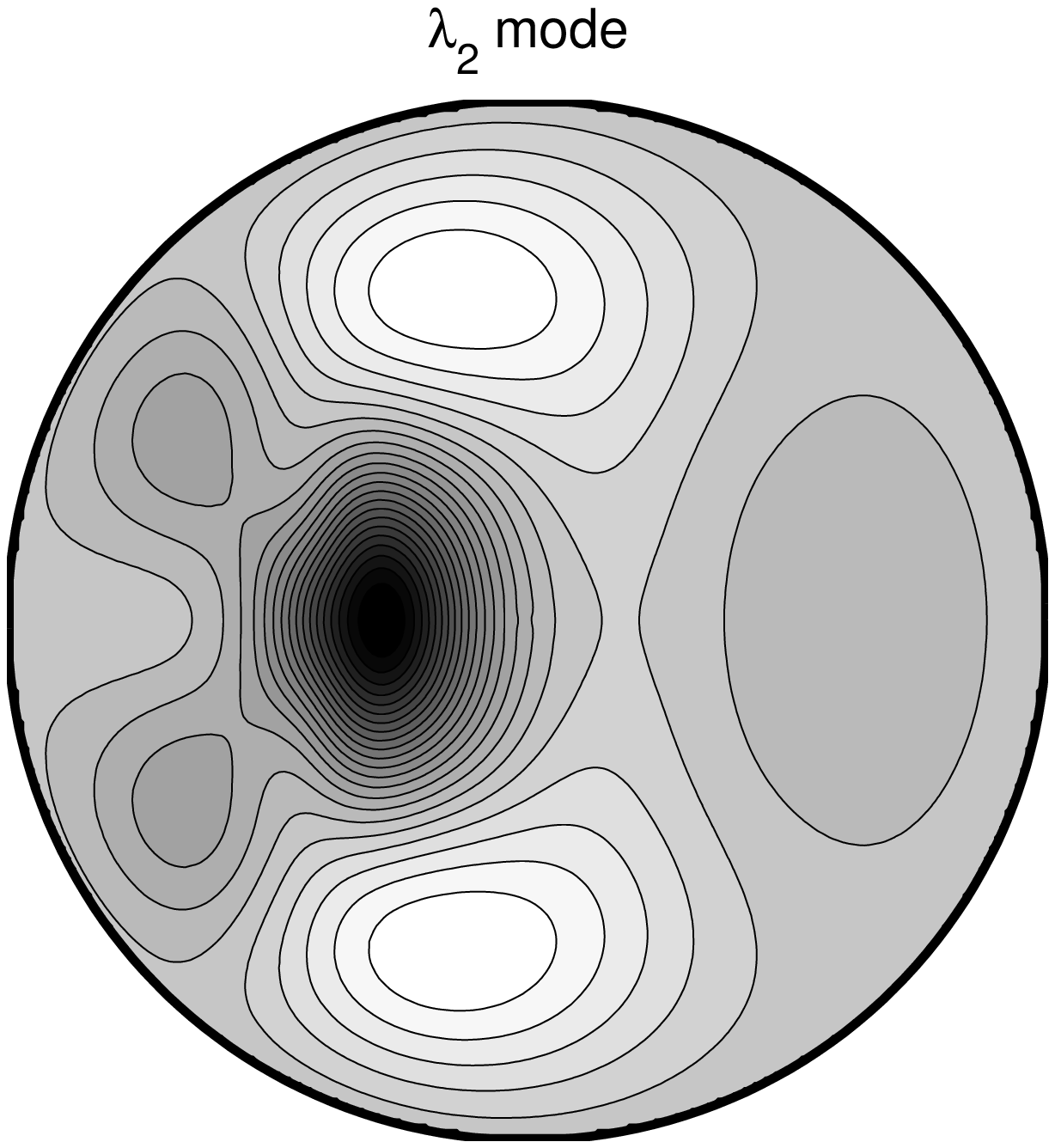}
\end{center}
\caption[xyz]{Contour plots of the $z$-averaged streamwise velocity
for the two unstable eigenvectors at $Re=2500$.  If the eigenvectors are
normalized to have unit kinetic energy, the level curves are
equispaced in the intervals $[-1.38,1.66]$ and $[-1.01,0.34]$,
respectively.  The lighter regions correspond to higher values.  At
other values of $Re$, the signs of the eigenvectors are chosen to yield
plots similar to the ones above.}
\label{fig-3}
\end{figure}

Figure \ref{fig-3} shows that the $\lambda_1$ eigenvector weakens the
high speed streaks of the traveling wave.  The effect of adding the
$\lambda_2$ eigenvector to the traveling wave would be to displace the
high speed streaks to a more symmetrical position.  It must be noted,
however, that the plot of the streaks of the $\lambda_2$ eigenvector is
not as meaningful because the $n=1$ mode is dominant. The two plots in
Figure \ref{fig-3} are used to assign positive and negative signs to
the eigenvectors at different $Re$ in a consistent manner. When
comparing cross-sections of velocity fields to traveling wave
solutions \citep{ES,HD,WillKer08}, it may be worthwhile to look at the
unstable eigenvectors of the traveling waves. The inevitable deviations
from the streak patterns of the traveling waves may correlate with the
streak patterns of the unstable eigenvectors.

\begin{figure}
\begin{center}
\includegraphics[scale=0.35]{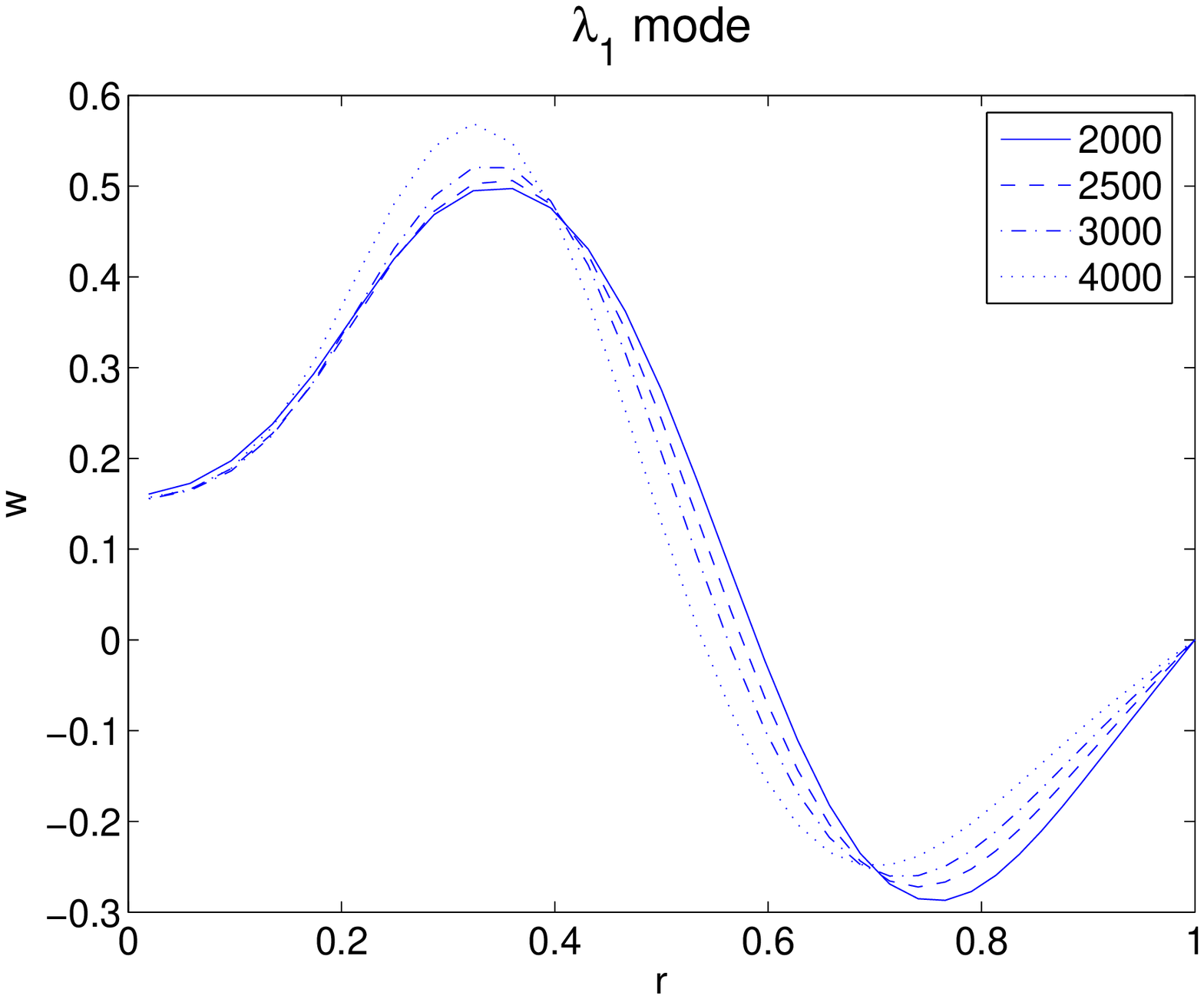}
\includegraphics[scale=0.35]{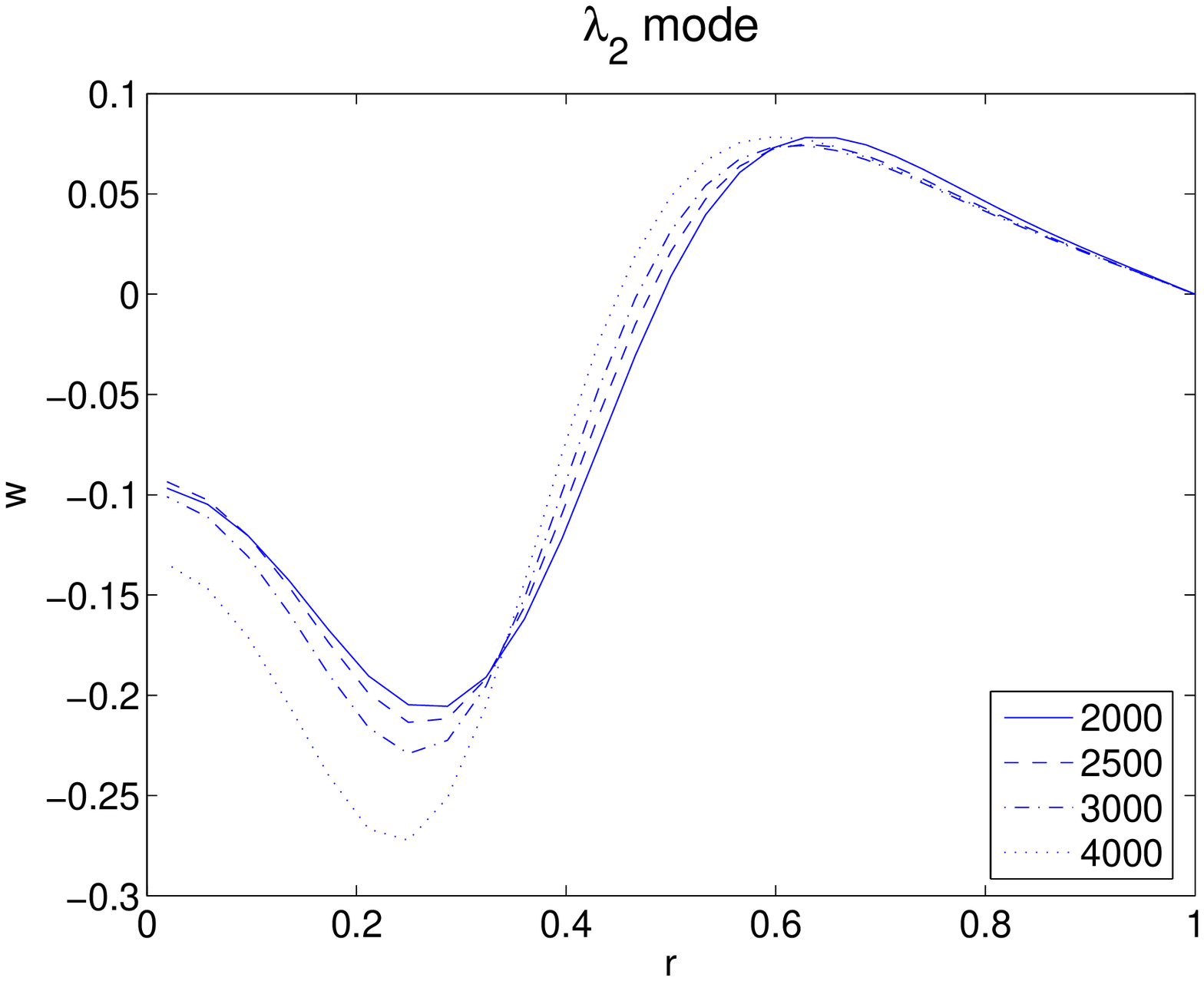}
\end{center}
\caption[xyz]{
Plots of the mean streamwise flow $\bar{w}(r)$ at various
values of $Re$.
The eigenvectors are
normalized to have kinetic energy equal to $1$.
}
\label{fig-4}
\end{figure}

Figure \ref{fig-4} shows the mean streamwise flow that corresponds to
the $\lambda_1$ and $\lambda_2$ eigenvectors. To form an idea of the
distortion to the mean flow of the laminar solution when those
eigenvectors are added as disturbances, the plots in Figure \ref{fig-4}
must be scaled by a factor of $1/50$ or less.

\begin{figure}
\begin{center}
(a)\includegraphics[scale=0.35]{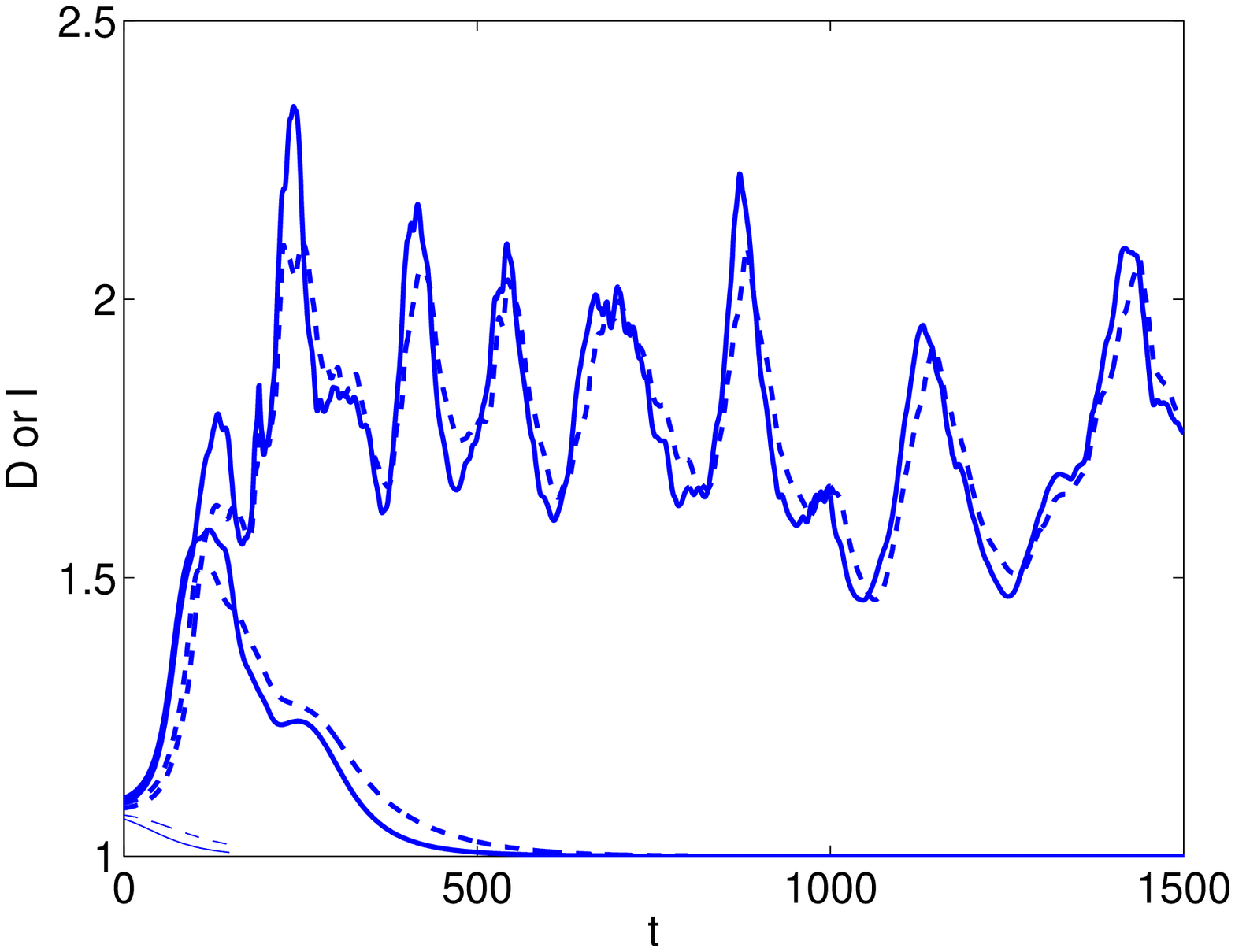}
(b)\includegraphics[scale=0.35]{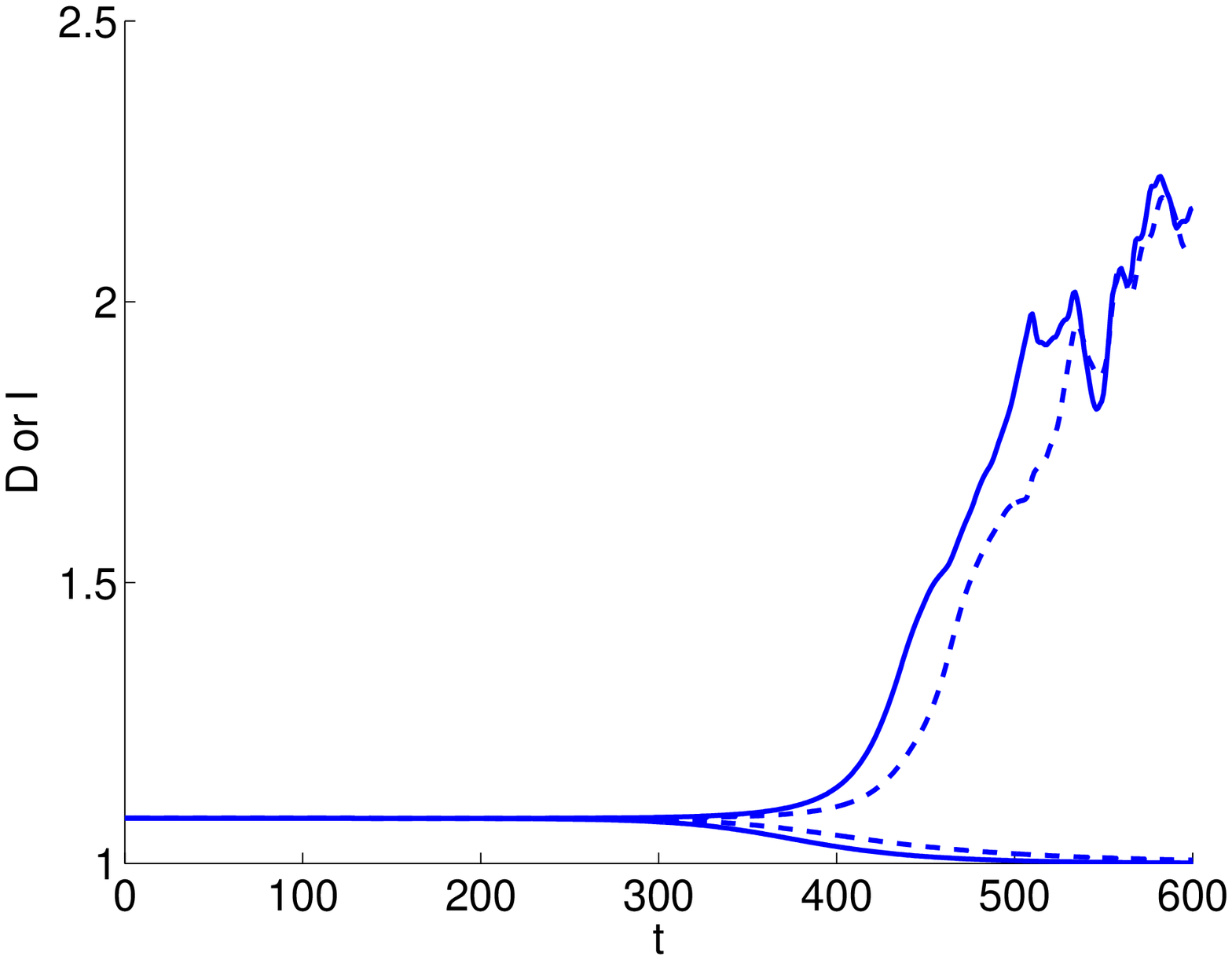}
\end{center}
\caption[xyz]{
(a) Plots of D (solid) and I (dashed) against time at $Re=2000$
and $Re=2500$. For both values of $Re$, perturbation of the traveling
wave by a positive multiple of the $\lambda_1$ eigenvector
shown in Figure
\ref{fig-4} leads to rapid laminarization (thin lines at lower left
corner). Perturbation by a negative multiple leads to a long transient
at $Re=2000$ and what appears to be sustained turbulence at
$Re=2500$. (b) Plots close to the edge.}
\label{fig-5}
\end{figure}

Figure \ref{fig-5}a shows that if the asymmetric traveling wave is
disturbed with a small and positive multiple of the $\lambda_1$
eigenvector, the disturbed state evolves and becomes laminar
uneventfully. That is unsurprising because the disturbance has the
effect of weakening the high speed streaks. In contrast, adding a
negative multiple leads to what appears to be sustained turbulence at
$Re=2500$. It is easily noticeable that energy dissipation $D$ is
greater than energy input $I$ when the plots in Figure \ref{fig-5}a
spike up, but is lesser when the plots dip down. Thus the kinetic
energy of the velocity field as a whole decreases during the spikes,
but increases during the dips. The decrease of kinetic energy during a
spike is  well correlated with flattening of the mean velocity
profile.

\begin{figure}
\begin{center}
(a) \psset{xunit=2cm, yunit=2cm}
\begin{pspicture}(-1.3,-1.7)(1,1)
\psline{->}(0,0)(1,0)
\rput[bl](1,.01){${\bf v_1}$}
\psline{->}(0,0)(0.71,0.71)
\psline{->}(0,0)(0,1)
\psline[linestyle=dashed]{->}(0,0)(-0.2, 0.98)
\rput[bl](-0.35,1.01){${\bf v_2}$}
\psline[linestyle=dashed]{->}(0,0)(-.71,.71)
\psline[linestyle=dashed]{->}(0,0)(-1,0)
\psline[linestyle=dashed]{->}(0,0)(-0.71,-0.71)
\psline[linestyle=dashed]{->}(0,0)(0,-1)
\psline[linestyle=dashed]{->}(0,0)(0.2,-0.98)
\psline(0,0)(0.71,-0.71)
\end{pspicture}
(b) \includegraphics[scale=0.4]{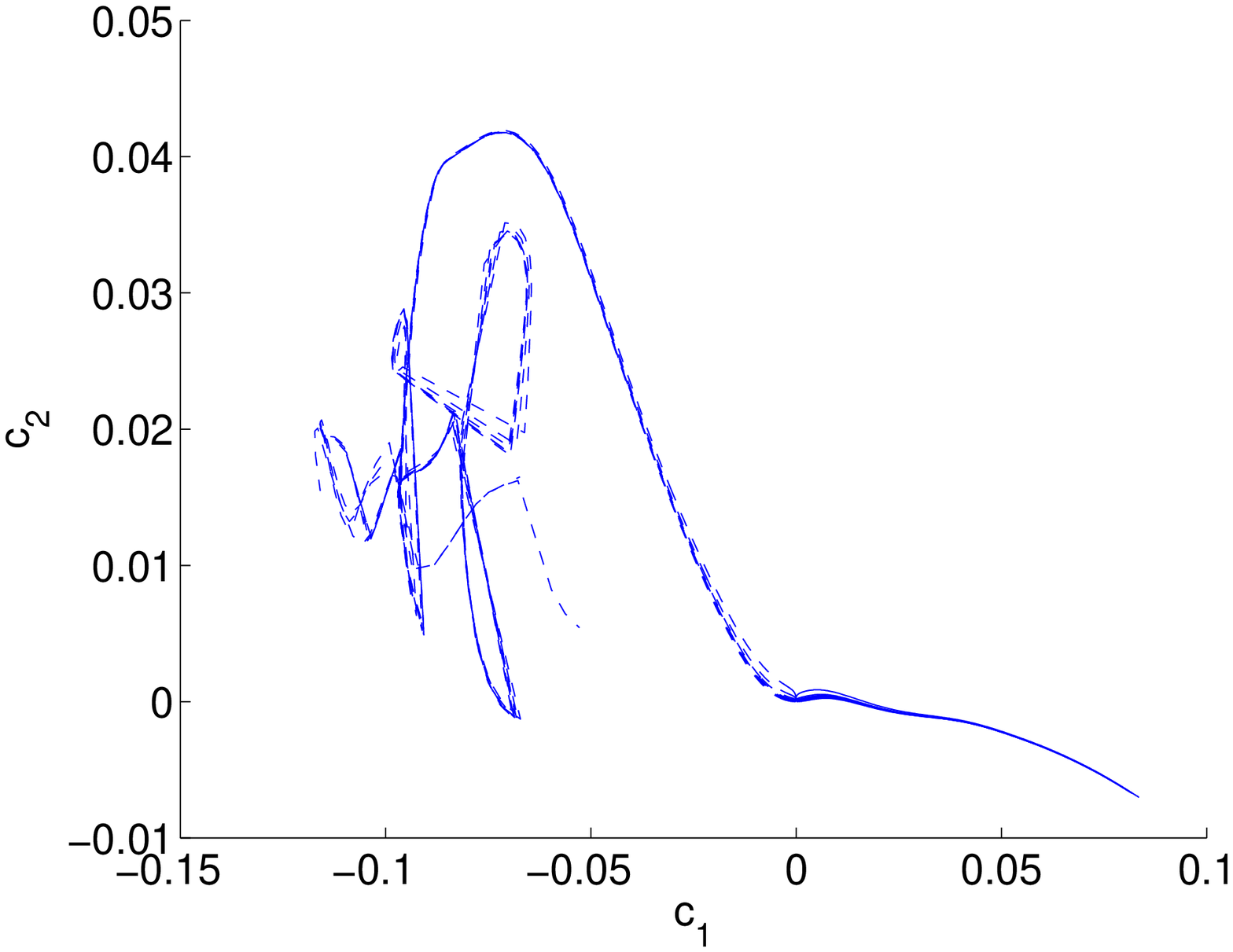}
\end{center}
\caption[xyz]{
(a) A schematic
sketch of directions on the unstable manifold,
where  the two eigenvectors are labeled. The
trajectories initiated in the dashed directions undergo turbulent
episodes.
(b) Projections of trajectories whose initial points are obtained
by perturbing the traveling wave at $Re=2500$ by a combination of the
$\lambda_1$ and $\lambda_2$ modes. The initial points are all near $(0,0)$.
The trajectories that laminarize turn right, while those that transition
to turbulence (dashed) turn left.
}
\label{fig-6}
\end{figure}

Figure \ref{fig-6}a shows a schematic sketch of the unstable
directions of the asymmetric traveling wave, while distinguishing
between directions that turn turbulent and ones that do not. Figure
\ref{fig-5}b corresponds to two trajectories close to the border
between turbulent and laminar directions. In that figure, the
trajectories near the border separate after $t=300$. By refining the
border, it appears that the point of separation can be deferred
indefinitely with a view to locating edge states.

To better visualize the unstable manifold, we adopt a technique
introduced by \cite{GHC} with the aim of getting good phase space
visualizations of turbulent trajectories. The
velocity field is a point in phase space and the evolution of an
initial velocity field with respect to the incompressible
Navier-Stokes equation is a trajectory in that phase space.  Because
the phase space is infinite-dimensional and does not lend itself to
plots directly, one has to use projections. An obvious projection
would be to pick some Fourier-Chebyshev modes from the discretization
of the velocity field. Although such projections have been employed,
they have a number of shortcomings. The choices of the component of
the velocity vector and of the mode of that component are both
arbitrary.  The component and the mode that are chosen capture only a
partial aspect of the velocity field. As a result of these
shortcomings, such projections look messy and one cannot form a good
idea of the dynamical structures in phase space from such projections.

Following \citep{GHC}, the projection we use picks a set of velocity
fields that appears well suited to visualize trajectories on the
unstable manifold.  Let ${\bf u}_{TW}$ be the velocity field of the
traveling wave and let ${\bf u}'$ and ${\bf u}''$ be an orthonormal
basis for its unstable space. The notion of orthogonality between
velocity fields corresponds to the kinetic energy norm.  We will
choose ${\bf u}'$ to be the same direction as the leading
eigenvector. With that choice the second eigenvector at $Re=2500$ is
approximately $ -0.13 {\bf u}' + 0.99 {\bf u}''$.  For each velocity
field that satisfies the shift-reflect symmetry
\eqref{eqn-2-5}, we obtain a projection in terms of ${\bf u}_{TW}$,
${\bf u}'$, and ${\bf u}''$. The velocity fields ${\bf u}'$ and ${\bf
u}''$ satisfy the no-slip boundary condition and have zero mass-flux.

Given a velocity field ${\bf u}$ that satisfies the shift-reflect
symmetry, one can decompose it as ${\bf u} - {\bf u}_{TW} = c_0 {\bf
u}' + c_1 {\bf u}'' + {\bf r},$ where the remainder ${\bf r}$ is
orthogonal to the plane of the eigenvectors.  One could use $c_0$ and
$c_1$ to represent ${\bf u}$ in a plot, but that would be
unsatisfactory. The problem is that one can translate ${\bf u}_{TW}$ in the
streamwise direction and obtain different velocity fields that stand
for the same wave. In order to eliminate dependence on translations
in the $z$ direction, we shift the velocity field ${\bf u}$ by $s_z$
in the $z$ direction and consider
\begin{equation}
{\bf u}(r, \theta, z+s_z) - {\bf
u}_{TW} = c_0 {\bf u}' + c_1 {\bf u}'' + {\bf r}_{s_z}.
\label{eqn-3-1}
\end{equation}
The shift $s_z$ is chosen to minimize $\norm{{\bf r}_{s_z}}$, and the
axes of the projection, $c_0$ and $c_1$, are the coefficients for that
shift.  The need to pick a shift $s_z$ arises because the equations of
pipe-flow are unchanged by translations along $z$.  The shift-reflect
symmetry is broken by rotations in the $\theta$ direction.  Since we
have restricted ourselves to vector fields with the shift-reflect
symmetry, shifts in $\theta$ are not considered in \eqref{eqn-3-1}.
The need to factor out
continuous symmetries arises in
ODEs \citep{GL-Gil07b}
and PDEs
such as the Kuramoto-Sivashinky equation as well \citep{CDS}.

Figure \ref{fig-6}b shows trajectories on
the unstable manifold using such a projection.
The initial  velocity fields were of the form
${\bf u}_{TW} + \epsilon a {\bf u}' + \epsilon b {\bf u}''$,
with $\epsilon = \times 10^{-4}$ and $a,b$ being scalars.
In all, we considered ten velocity fields corresponding to
$(a,b)=(\pm 1, \pm 1)$, the four coordinate directions,
and two directions along the second eigenvector. Since
$\epsilon$ is small, all these velocity fields were
very nearly on the unstable manifold. The distinction between
trajectories that laminarize uneventfully and those that
undergo a turbulent episode is  clear in Figure \ref{fig-6}b.


In the next section, we return to such projections of the
unstable manifold to partially justify arguments used to find
disturbances of the laminar flow which evolve and hit the asymmetric
traveling wave.

\section{Hitting the traveling wave at $Re=2000$ and $Re=2500$}

\begin{table}
\begin{center}
\scriptsize
\begin{tabular}{ccccccccc}
$f_r$ & $f_1$& $f_2$ & $T$ & $\delta$ & $ke_0$ & $ke_1$ & $ke_2$ & $ke_3$ \\ \hline
$7.084220e\!\!-\!\!3$ & $-2.114944e\!\!-\!\!2$ & $0$ & $371.49$ & $8.1e\!\!-\!\!4$ & $9.8e\!\!-\!\!1$ & $4.3e\!\!-\!\!4$ & $2.2e\!\!-\!\!5$ & $2.1e\!\!-\!\!7$ \\ 
$1.242439e\!\!-\!\!2$ & $1.890000e\!\!-\!\!2$ & $0$ & $219.70$ & $8.9e\!\!-\!\!3$ & $9.8e\!\!-\!\!1$ & $3.3e\!\!-\!\!4$ & $1.4e\!\!-\!\!5$ & $1.0e\!\!-\!\!7$ \\ 
$9.119378e\!\!-\!\!3$ & $0$ & $1.720663e\!\!-\!\!2$ & $257.80$ & $4.2e\!\!-\!\!3$ & $9.8e\!\!-\!\!1$ & $4.1e\!\!-\!\!4$ & $2.0e\!\!-\!\!5$ & $1.9e\!\!-\!\!7$ \\ 
$1.017473e\!\!-\!\!2$ & $0$ & $-1.400000e\!\!-\!\!2$ & $240.90$ & $5.5e\!\!-\!\!3$ & $9.8e\!\!-\!\!1$ & $3.7e\!\!-\!\!4$ & $1.7e\!\!-\!\!5$ & $1.4e\!\!-\!\!7$ \\ 
$9.048182e\!\!-\!\!3$ & $3.923814e\!\!-\!\!4$ & $1.757170e\!\!-\!\!2$ & $281.40$ & $2.9e\!\!-\!\!3$ & $9.8e\!\!-\!\!1$ & $4.2e\!\!-\!\!4$ & $2.2e\!\!-\!\!5$ & $2.0e\!\!-\!\!7$ \\ \hline\hline
$5.687034e\!\!-\!\!3$ & $-1.643322e\!\!-\!\!2$ & $0$ & $323.84$ & $2.9e\!\!-\!\!3$ & $9.8e\!\!-\!\!1$ & $2.5e\!\!-\!\!4$ & $1.0e\!\!-\!\!5$ & $7.7e\!\!-\!\!8$ \\ 
$9.591329e\!\!-\!\!3$ & $1.562075e\!\!-\!\!2$ & $0$ & $318.60$ & $7.7e\!\!-\!\!3$ & $9.8e\!\!-\!\!1$ & $3.1e\!\!-\!\!4$ & $1.5e\!\!-\!\!5$ & $1.4e\!\!-\!\!7$ \\ 
$7.049209e\!\!-\!\!3$ & $0$ & $1.420195e\!\!-\!\!2$ & $268.00$ & $4.8e\!\!-\!\!3$ & $9.8e\!\!-\!\!1$ & $2.4e\!\!-\!\!4$ & $9.4e\!\!-\!\!6$ & $6.9e\!\!-\!\!8$ \\ 
$7.986395e\!\!-\!\!3$ & $0$ & $-1.115761e\!\!-\!\!2$ & $309.80$ & $6.0e\!\!-\!\!3$ & $9.8e\!\!-\!\!1$ & $2.2e\!\!-\!\!4$ & $8.1e\!\!-\!\!6$ & $5.5e\!\!-\!\!8$ \\ 
$5.676511e\!\!-\!\!3$ & $-1.646355e\!\!-\!\!2$ & $8.768074e\!\!-\!\!5$ & $292.70$ & $3.7e\!\!-\!\!3$ & $9.8e\!\!-\!\!1$ & $2.9e\!\!-\!\!4$ & $1.3e\!\!-\!\!5$ & $1.2e\!\!-\!\!7$
\end{tabular}
\normalsize
\end{center}
\caption[xyz]{
Data at $Re=2000$ (above the double line) and at $Re=2500$
(below the double line). The $f$s are magnitudes of disturbances of
the laminar solution. $T$ is the time of closest approach to the
traveling wave. $\delta$ is the distance from the traveling wave at that
time. The last four columns give the kinetic energies in the $n=0, \pm
1, \pm 2, \pm 3$ modes at the point of closest approach.}
\label{table-4}
\end{table}

The $f_r$ column of Table \ref{table-4} gives the norm of the rolls.
The velocity field of the rolls, denoted by ${\bf u}_r$,
is obtained by averaging the traveling wave in the streamwise
direction and discarding the streamwise component of the velocity.
The norms of the unstable eigenvectors are denoted by $f_1$ and $f_2$.
The velocity fields of the eigenvectors are denoted by ${\bf u}_1$
and ${\bf u}_2$. The laminar solution is ${\bf u}_L$ and the traveling
wave is ${\bf u}_{TW}$.

The distance $\delta$ of closest approach listed in Table \ref{table-4} is
obtained as follows. The Navier-Stokes equation is integrated from
the initial velocity field ${\bf u}_L + f_r {\bf u}_r + f_1 {\bf u}_1
+ f_2 {\bf u}_2$.
The initial velocity field has the shift-reflect symmetry and so does
${\bf u}_t$,
where ${\bf u}_t$ is the velocity field at time $t$.
Define
\begin{equation}
\delta(f_r, f_1, f_2) = \min_{t\geq 0} \min_{0\leq s_z < 2\pi\Lambda} \norm{
{\bf u}_t(r,\theta, z+s_z) - {\bf u}_{TW}}.
\label{eqn-4-1}
\end{equation}
To compare ${\bf u}_t$ and ${\bf u}_{TW}$, one has to minimize over
shifts $s_z$ for the same reason as in \eqref{eqn-3-1}. To find the
minimizing shift $s_z$, we first try $s_z = \pi\Lambda k/N$, $0\leq k
< 2N$. Using that data, an interval that contains the minimum is
found and that interval is refined recursively to a depth equal to $30$.
We refer to the result of the inner minimum in \eqref{eqn-4-1} as the
distance between ${\bf u}_{TW}$ and ${\bf u}_t$. This method of
finding that distance is expensive, with the cost of finding the
distance being more than $20$ times the cost of a single time
step. However, it finds the distance with an accuracy of $4$ or $5$
digits.

Given the expense of finding the distance between ${\bf u}_t$ and
${\bf u}_{TW}$, the distance being the inner minimum in
\eqref{eqn-4-1}, care has to be exercised in finding the outer
minimum over $t$. If the distance is computed after every time step,
the cost of the computation becomes prohibitive. The wall time for
integrating a velocity field for a time interval of $100$ is about an
hour on an Opteron processor, but becomes more than $20$ hours if 
the distance to $\bf{u}_{TW}$ is computed after every time step. For an
initial waiting time when the streaks are still forming, we do not
compute the distance at all. This waiting time is longer for larger
$Re$.  Thereafter the distance is computed every $100$ time steps
only, a time step being $0.01$. As the distances vary smoothly as a
function of time, we use polynomial extrapolation to predict if the
distance function has a minimum within the next $100$ time steps or
not. If it is predicted to have a minimum within the next $100$ time
steps, we measure the distance every $10$ time steps. If the distance
function is predicted to have a minimum within the next $10$ time
steps, we measure the distance after every time step. The value of
$\delta$ is the first local minimum found in this manner and it is
very probably also the global minimum over $t$. The time step can be
successively decreased to get finer estimates of $\delta$, but that
was not implemented.

\begin{figure}
\begin{center}
\includegraphics[scale=0.4]{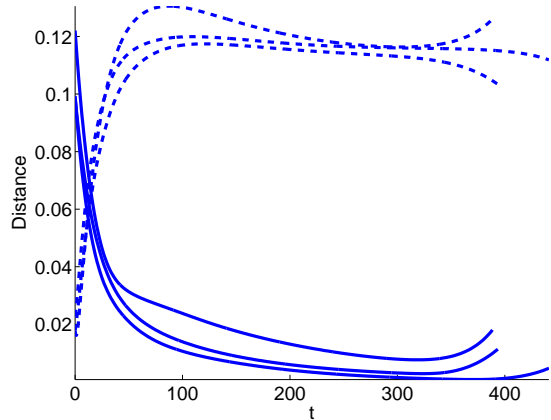}
\end{center}
\caption[xyz]{
Distances from the laminar solution (dashed) and the traveling wave (solid).
The three pairs of curves correspond to the first line with $Re=2000$
in Table \ref{table-4} and the first two lines with $Re=2500$. Data in that
table may be used to deduce the assignment of curves above to rows in that
table.}
\label{fig-7}
\end{figure}

The times $T$ at which the minima were attained are given in Table
\ref{table-4}. $T$ is measured with a precision of $.01$ in only two
lines of that table. The measurements in the other lines have a
precision of $0.1$. If the last four columns of that table are
compared with the last four columns of Table \ref{table-1}, the
comparison confirms that the approach to the traveling wave is closer
when $\delta$ is smaller. Figure \ref{fig-7} leaves no room for doubt
that the first disturbances listed for $Re=2000$ and $Re=2500$ in
Table \ref{table-4} evolve and hit the corresponding traveling waves.
The smallest distance $\delta$ from the traveling wave is realized
after time $T$, which is listed in Table \ref{table-4}.  Figure
\ref{fig-7} shows that the plots of the distances from the traveling
wave and the laminar solution both become flat around $t=T$.  The
disturbance moves away from the laminar solution rapidly at $t=0$. In
contrast, for heteroclinic connections there are two flat regions in
$t$, which correspond to time spent in the neighborhoods of the
invariant solutions joined by the heteroclinic connection
\citep{GHCV08}.

We are yet to explain the method used to find the numbers $f_r$,
$f_1$, and $f_2$ in Table \ref{table-4}. With each of those
disturbances to the laminar solution, the disturbed state lands close
to the stable manifold of the traveling wave and evolves to make a
close approach of small $\delta$ to the traveling wave. Each line
Table \ref{table-4} was obtained by minimizing $\delta(f_r, f_1, f_2)$
in different ways. The manner of minimization will now be described.

Although $\delta$ has three arguments corresponding to three
disturbances, each minimization was two dimensional.  The unstable
manifold of the traveling wave at the $Re$ under consideration is two
dimensional as shown by Table \ref{table-2}. If the directions that
correspond to translating and rotating the traveling wave are ignored,
the co-dimension of the traveling wave is two. The inner minimization
in \eqref{eqn-4-1} accounts for streamwise translations, while
rotations around the pipe axis would break the shift-reflect
symmetry. Thus the stable manifold is in effect a co-dimension two
object.

Suppose there is some system of coordinates for the
infinite-dimensional phase space in which the traveling wave is
$(0,0,0,\ldots)$. Suppose further that its unstable manifold is given
by fixing all except the first two coordinates at zero and its stable
manifold is obtained by fixing the first two coordinates at zero.  To
disturb the laminar solution on to the stable manifold, the first two
components must be zeroed out. Generically, it is impossible to zero
out two components by varying the amplitude of a single
disturbance. That is why we varied two disturbances.

All the rows of Table \ref{table-4} have $f_r > 0$. Adding the rolls
to the laminar solution causes the flow to develop streaks of
approximately the right form.  But the key to hitting the traveling
wave is to disturb the laminar state in such a way that the evolving
velocity field is free from unstable directions as it approaches the
traveling wave. If we were allowed to make disturbances near the
traveling wave, we could simply eliminate the unstable
directions. Since the disturbances are made to the laminar solution,
we have to somehow guess the directions at $t=0$ which evolve into
unstable directions at $t=T$, the time of closest approach. Analysis
can possibly suggest a better choice, but we simply used the unstable
directions ${\bf u}_1$ and ${\bf u}_2$ to disturb the laminar flow. In
addition the magnitude of the rolls themselves can be varied to get a
direction that has a nonzero component along the unstable manifold at
$t=T$.

\begin{figure}
\begin{center}
\includegraphics[scale=0.4]{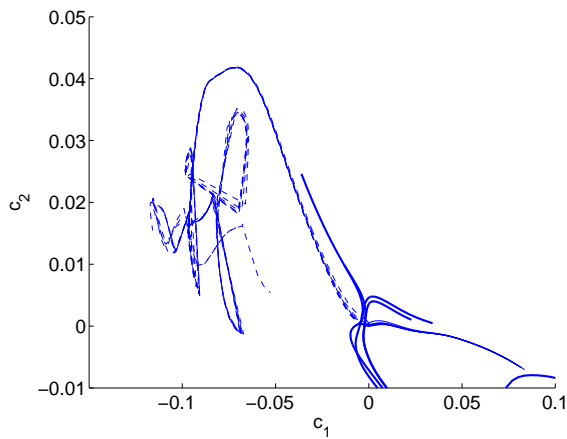}
\end{center}
\caption[xyz]{
Similar to Figure \ref{fig-6}a, but the thick lines show the projections
of trajectories at $Re=2500$
which are initialized using the disturbances in Table \ref{table-4}.
}
\label{fig-8}
\end{figure}

Each row of Table \ref{table-4} that has $f_1 = 0$ or $f_2 = 0$ was
gotten by fixing that disturbance at $0$ and minimizing over the other
two disturbances. In addition, the sign of the other disturbance that
adds an eigenvector was prescribed. Thus there are four rows of that
type for $Re=2000$ and $Re=2500$. For the last row with $Re=2000$ or
$Re=2500$, $f_r$ was fixed while $f_1$ and $f_2$ were varied.  Figure
\ref{fig-8} shows that for disturbances at $Re=2500$, the flow evolves
to a state where its dynamics is governed mainly by the unstable
manifold of the traveling wave, thus partially supporting the
reasoning used to find those disturbances.

\begin{figure}
\begin{center}
\includegraphics[scale=0.4]{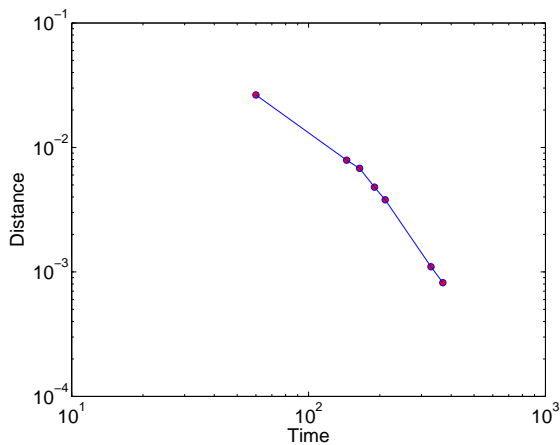}
\end{center}
\caption[xyz]{
The axes correspond to the $T$ and $\delta$ columns in Table \ref{table-4}.
The plot is for $Re=2000$ and shows the progress of a 
single minimization (first line of Table \ref{table-4}). Each point corresponds to a certain stage
in the sequence of optimizations used to find a disturbance of the
laminar flow such that the disturbed flow evolves and hits the
traveling wave.  }
\label{fig-9}
\end{figure}

When $\delta(f_r, f_1, f_2)$ is minimized numerically, the
disturbances found at successive stages of the minimization give
smaller $\delta$ but with larger values of $T$, the time of closest
approach to the traveling wave, as shown in Figure \ref{fig-9}. For
the theoretical ideal $\delta = 0$, $T$ would be infinite. Thus the
numerical optimization becomes progressively more expensive.

A more severe impediment to numerical minimization is the non-smooth
dependence of $\delta$ on the disturbances when $\delta = 0$. Because
the time to hit the traveling wave diverges, even a small change in the
disturbances causes a big change in the value of $\delta$.

The numerical optimization was implemented using Matlab's $fmincon()$,
which allows constraints to be placed on the values of $f_r$ or $f_1$
or $f_2$. The C++ code for computing the function $\delta(f_r, f_1,
f_2)$ was invoked from Matlab. The unconstrained version $fminunc()$
was not used because it tends to take such large steps while varying
$f_r$ or $f_1$ or $f_2$ that the numerical integration of the
Navier-Stokes equation becomes unstable. Because of the
non-smoothness, a nonlinear least squares solver, such as Matlab's
$lsqnonlin()$, might be a better option than
$fmincon()$. $lsqnonlin()$ minimizes $\sqrt{\abs{x+2}}$ from $x=3$
with just $6$ function evaluations while $fmincon()$ takes $63$
function evaluations to find a slightly worse approximation to the
minimum.  It was not used, however, because it does not provide the
facility to constrain the arguments.

The choice of initial guesses for the disturbances is not much of an
issue because the numerical optimization is relatively efficient at
the early stages. However, as $\delta=0$ is approached, the
optimization routine tries unrealistically large steps,
necessitating a lot of wasteful backtracking. It is difficult to assess
the quality of the search directions. Each row in Table \ref{table-4}
required at least $200$ hours of computing and often significantly
more. The first rows with $Re=2000$ and $Re=2500$ required much more
than $1000$ hours to attain smaller values of $\delta$. The
computations required repeated manual intervention to reset the
parameters to $fmincon()$, which is the reason we were able to
run the numerical optimization longer for only two rows of
Table \ref{table-4}.

\begin{table}
\begin{center}
\scriptsize
\begin{tabular}{c|ccccccccc}
$Re$ & $f_r$ & $f_1$& $f_2$ & $T$ & $\delta$ & $ke_0$ & $ke_1$ & $ke_2$ & $ke_3$ \\ \hline
$2000$ & $9.119378e\!\!-\!\!3$ & $0$ & $1.720663e\!\!-\!\!2$ & $257.80$ & $4.2e\!\!-\!\!3$ & $9.8e\!\!-\!\!1$ & $4.1e\!\!-\!\!4$ & $2.0e\!\!-\!\!5$ & $1.9e\!\!-\!\!7$ \\ 
$2500$ & $7.049209e\!\!-\!\!3$ & $0$ & $1.420195e\!\!-\!\!2$ & $268.00$ & $4.8e\!\!-\!\!3$ & $9.8e\!\!-\!\!1$ & $2.4e\!\!-\!\!4$ & $9.4e\!\!-\!\!6$ & $6.9e\!\!-\!\!8$ \\ 
$3000$ & $6.466047e\!\!-\!\!3$ & $0$ & $-9.890996e\!\!-\!\!3$ & $356.61$ & $7.0e\!\!-\!\!3$ & $9.8e\!\!-\!\!1$ & $2.2e\!\!-\!\!4$ & $9.4e\!\!-\!\!6$ & $8.4e\!\!-\!\!8$ \\
$4000$ & $4.600000e\!\!-\!\!3$ & $0$ & $8.829559e\!\!-\!\!3$ & $386.95$ & $9.5e\!\!-\!\!3$ & $9.8e\!\!-\!\!1$ & $6.9e\!\!-\!\!5$ & $1.5e\!\!-\!\!6$ & $7.1e\!\!-\!\!9$
\end{tabular}
\normalsize
\end{center}
\caption[xyz]{
Data at various $Re$. The columns are as in Table \ref{table-4}.}
\label{table-5}
\end{table}

Table \ref{table-5} shows that the magnitude of the disturbances of
the laminar flow required to hit the traveling wave diminishes with
$Re$. The quality of the approach to the traveling wave degrades with
increasing $Re$.  The quality of the approach can be assessed using
the $\delta$ column of Table \ref{table-5} and by comparing the last
four columns of that table to the last four columns of Table
\ref{table-1}.  We are not certain why the numerical minimization
has worse performance for increasing $Re$, although it could be
because the non-smoothness issue gets worse as $Re$ increases. The
tendency of the eigenvalues to approach the imaginary axis at varying
rates as $Re$ is increased may have something to do with it.

\section{Conclusion}

The aim of this paper was to find a method to disturb the laminar flow
such that the disturbed state evolves and hits a given traveling
wave. For the asymmetric traveling wave of \cite{PK}, we showed that
certain linear combinations of rolls and the two unstable directions
generate disturbances of the laminar solution that evolve and hit the
traveling wave. Numerical minimization was used to find linear combinations
that achieve that effect. As the numerical minimization comes closer
and closer to finding disturbances that evolve and hit the traveling wave,
the minimization problem becomes non-smooth.

It is reasonable to conjecture that this method of disturbing the
laminar flow so that it evolves and hits a given traveling wave is
applicable to other lower branch solutions of pipe flow and the
channel flows. However, the computational effort will increase with
the number of unstable directions. As indicated in the text, it may
be possible to use analysis to find disturbances that work better than
the unstable eigenvectors of the traveling wave.

It is yet unknown if spatially localized structures such as puffs in
transitional pipe flow and turbulent spots in plane Couette flow
correspond in their entirety to invariant solutions of the Navier-Stokes
equation. The existence of such invariant solutions is a topic
worthy of investigation. If such solutions are indeed found, the logic
used to find disturbances that evolve and hit the asymmetric
traveling wave will become applicable to transition in pipes of
realistic lengths with consequences for experimental investigations.

\noindent{\bf Acknowledgments.}
The authors thank J.F. Gibson and the referees for helpful
discussions.  The Center for Advanced Computing at the University of
Michigan provided computing facilities.  DV was partly supported by
NSF grants DMS-0407110 and DMS-0715510.  PC was partly supported by
NSF grant DMS-0807574.

\bibliography{references}
\bibliographystyle{plainnat} \end{document}